\documentclass[aps,prd,preprint,tightenlines,superscriptaddress,nofootinbib]{revtex4}
\usepackage{amsmath,amssymb,url}
\usepackage{graphicx,tabularx}

\def\bq{\begin{equation}}
\def\eq{\end{equation}}
\def\ba{\begin{eqnarray}}
\def\ea{\end{eqnarray}}

\newcommand{\etal}{\textit{et al.}}

\newcommand{\eg}{{\sl e.g. }}

\newcommand{\aj}{\gamma j}
\newcommand{\as}{\alpha_s}

\newcommand{\aem}{\alpha_{\rm EM}}

\begin{document}

\preprint{}

\date{\today}

\title{Testing Grand Unification at the (S)LHC}

\author{D.~Rainwater}
\email{rain@pas.rochester.edu}
\affiliation{Dept.~of Physics and Astronomy,
             University of Rochester, Rochester, NY, USA}
\author{T.M.P.~Tait}
\email{tait@anl.gov}
\affiliation{High Energy Physics Division, Argonne National Lab,
             Argonne, IL, USA}

\begin{abstract}
We examine the possibility of measuring the three gauge couplings at
high scales at the LHC, in order to see the first steps as they run
toward Grand Unification at much higher energies.  Using the MSSM with
sparticle masses of several hundred GeV as an example of a theory in
which the couplings do unify at very high energies, we find that the
processes $pp\to\ell^+\nu$, $pp\to\ell^+\ell^-$ and $pp\to\gamma j$
can be useful to discriminate
the SM from the MSSM with masses at the few hundred GeV scale, and
determine that the couplings are converging at better than the SM
prediction toward the GUT scale.  Such measurements indirectly probe the
existence of lower mass states, charged under the SM gauge groups, but
which may be difficult to produce directly or extract from backgrounds
at the LHC.
\end{abstract}

\maketitle

%%%%%%%%%%%%%%%%%%%%%%%%%%%%%%%%%%%%%%%%%%%%%%%%%%%%%%%%%%%%%%%%%%%%%%%%
%%%%%%%%%%%%%%%%%%%%%%%%%%%%%%%%%%%%%%%%%%%%%%%%%%%%%%%%%%%%%%%%%%%%%%%%

\section{Introduction}

The Standard Model (SM) of particle physics is an extremely successful
description of nature at the subnuclear level, although it is also
generally regarded as an effective theory, likely not valid beyond the
TeV scale.  For example, the exact mechanism of electroweak symmetry
breaking remains undetermined.  The SM explanation contains a minimal
weakly-coupled Higgs sector, which introduces a number of theoretial
loose ends, including the gauge hierarchy problem, and the lack of a
compelling theory to determine the pattern of flavor.  In addition,
there is strong evidence for the existence of large amounts of
non-baryonic ``dark'' matter in the universe, which the SM does not
include.  These and other puzzles have led to the development of
numerous extensions to the SM, varying greatly in their underlying
structure and new particle content.  A common theme, however, is the
presence of new heavy states in the sub-Tev to few-TeV mass range.
The LHC at CERN will soon study the TeV mass scale, and is expected to
probe these mysteries at a level previously unobtainable in particle
physics.

One of the primary goals in exploring higher energy scales is the hope
that simpler organizing principles can be revealed and understood.  In
the same way that QCD took the effective theory of mesons and baryons,
described by a huge number of states each with its own mass and
couplings, and allowed (at least) a qualitative understanding of the
pattern in terms of a gauge theory with a small handful of parameters,
it is hoped that at higher energies we will understand how to predict
the structure and parameters of the SM in terms of simpler principles
and fewer input parameters.  Chief among these potential triumphs is
the idea that we will be able to understand the $SU(3)_C\times
SU(2)_W\times U(1)_Y$ gauge structure of the SM (and the
representations of the fermions under it) in terms of a single grand
unified theory (GUT) with a single coupling constant.  The minimal
simple group containing the SM gauge structure, $SU(5)$, has some
remarkable successes: na\"ive extrapolation of the SM gauge couplings
does seem to move toward convergence as one considers physics at
smaller distances, and the known SM matter fields precisely fill out
complete $SU(5)$ representations.

\begin{figure}[ht!]
\begin{center}
\includegraphics[scale=0.5]{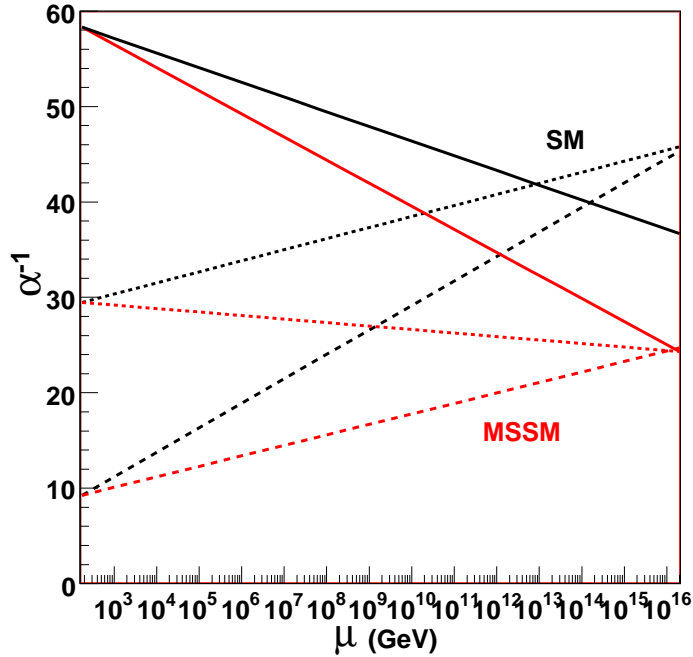}
\end{center}
\vspace{-10mm}
\caption{Evolution of the three gauge couplings $\alpha_1$ ($SU(5)$ 
normalized, as the dashed line), $\alpha_2$ (dotted line) and
$\alpha_3$ (solid line) in the SM (upper set of curves) and MSSM
(lower set of curves), assuming superpartner masses are all close to
the top quark mass.}
\label{fig:gut}
\end{figure}

However, despite these encouragements, grand unification in the SM
does not quite work.  The couplings do converge, but they fail to meet
one another at a single scale, as shown in Fig.~\ref{fig:gut}.  The
amount of discrepancy is large ($\sim 20\%$) compared with any
expected correction from thresholds of heavy fields at the GUT scale
itself.  Thus, in order for the couplings to actually meet at high
energies, one must include some correction associated with a lower
energy scale, to alter the running and induce a log-enhanced shift in
the coupling with respect to the SM prediction.  This argues that the
new states responsible for unification of the couplings should be
relatively light\footnote{However, see Ref.~\cite{Bourilkov:2006rz}, which
claims that very light supersymmetrc particles (in a simplistic
framework) are already indicated by comparison of couplings at the $Z$
pole with those at energies of order 200 GeV.}  (though it does not
guarantee that they will be within the reach of the LHC).

In particular, the minimal supersymmetric standard model (MSSM), is a
well-motivated extension of the SM, which stabilizes the electroweak
scale with respect to quantum corrections from higher scales, and
(usually) contains a viable dark matter candidate with an appropriate
relic density.  In addition, when one includes the full MSSM particle
content and extrapolates the gauge couplings to large energies, one
finds remarkable convergence\footnote{Unification at a similar level
also occurs by adding vector-like quarks \cite{Vlike-quarks}.}  at a
scale of order $10^{16}$~GeV (see Fig.~\ref{fig:gut}).  While previous
searches for supersymmetric particles have had null
results~\cite{run2}, it seems likely that if TeV-scale
supersymmetry is realized in nature, the LHC will be able to discover
it.  In particular, the LHC is expected to copiously produce the
colored superpartners in large numbers~\cite{Dawson:1983fw}.  The
discovery of superpartners would be a strong hint that the couplings
do unify, and that a GUT is realized in nature.

In this article we explore the possibility that the LHC could find at
least the first hints of Grand Unification, by measuring the gauge
couplings from observation of ordinary SM processes at high scales.
If new particles exist at the weak scale, they will influence the
evolution of these couplings at energies larger than their masses, and
the couplings will begin to run away from the SM extrapolations.
While the LHC has the potential to see only the first step in eventual
unification of the couplings, we will see that it can make a statement
as to whether the couplings seem to be coming together as in the MSSM,
or missing each other, as in the SM.  We will use the MSSM with masses
of several hundred GeV as our test example, though one of the virtues
of the approach is that it is not very sensitive to the underlying
model itself.

The corrections that cause the coupling to run differently amount to a
class of higher-order virtual corrections from the new states running
in the loops.  At energy scales of order the mass of the new
particles, such corrections are complicated, and not described
suficiently by simply running the coupling.  We envision that at the
highest energies the LHC probes (the multi-TeV region), we are
sufficiently above the new mass thresholds that the log-enhanced terms
dominate the corrections, and the description in terms of a running
coupling captures the bulk of the effect of the new physics.  Of
course, this implies that the LHC has enough energy to produce the new
states directly and would likely have seen at least some of them.
Even so, a direct extrapolation including the new states explicitly
may not be possible.  It is well known in the MSSM that while the
colored superpartners are produced in large numbers, the electroweakly
interacting sparticles will in most cases be visible only if they form
part of an observable decay chain from the colored objects; in most
regions of MSSM parameter space, some of the electroweak objects will
end up being missed.  Even when they are visible, the representations
under $SU(2)_W\times U(1)_Y$ are obscured by electroweak symmetry
breaking, and the representation content is likely to be impossible to
determine using LHC data alone.

We identify three processes at LHC which could potentially provide
such a measurement.  They naturally must be observable with good
statistics at large invariant mass, in particular above the mass
thresholds of SUSY particles, and with systematic uncertainties much
smaller than the change in rate due to altered coupling evolution.
The obvious candidates are: Drell-Yan (DY) production of mixed
charged-neutral lepton pairs (off-shell $W$ production), sensitive at
leading order (LO) only to the $SU(2)$ coupling; DY charged lepton
pairs (through off-shell $Z/\gamma^*$), sensitive at tree level to
both the $SU(2)$ and $U(1)_Y$ couplings; and photon-jet ($\aj$)
production, which is sensitive to both QCD's $SU(3)$ coupling as well
as the $U(1)_{EM}$ coupling, itself a combination of both the $SU(2)$ and
$U(1)_Y$ couplings: $\alpha_{EM}=\alpha_1\alpha_2/(\alpha_1+\alpha_2)$.

We review the running of gauge couplings and give our input parameters
in Sec.~\ref{sec:run}, then calculate the cross sections expected at
LHC for the three processes in Sec.~\ref{sec:xsecs}, along with our
estimates for how accurately the MSSM rates could be distinguised from
those of the SM.  Finally, we discuss the prospects and for performing
these measurements at LHC, and the uncertainties of greatest concern.

%%%%%%%%%%%%%%%%%%%%%%%%%%%%%%%%%%%%%%%%%%%%%%%%%%%%%%%%%%%%%%%%%%%%%%%%

\section{Running of the Gauge Couplings
\label{sec:run}}

In an effective theory approach, we capture the leading log
corrections from higher orders by employing a set of running
couplings,
\bq\label{eq:run}
\alpha_i(Q^2) \; = \;
\frac{\alpha_i(\mu^2)}{1-\beta_i\,\alpha_i(\mu^2)/ 4 \pi \ln(Q^2/\mu^2)}
\; ,
\eq
where $\alpha_i=g_i^2/4\pi$ is the analog of the fine structure
constant for each of the SM's three gauge groups ($U(1)_Y$, $SU(2)$,
and $SU(3)$), $\mu$ is a reference scale at which the coupling has
already been measured (usually $M_Z$ is a convenient choice) and $Q$
is the scale of the physical process in question.  The coefficients
$\beta_i$ ($i=\{ 1,2,3 \}$ for $\{ U(1)_Y, SU(2)_W, SU(3)_C\}$)
are determined by the particle content active at scales of
order $Q$: those species whose masses are less than $Q$.  Mass
thresholds appear (at this order) as changes in $\beta_i$, resulting
in a change in the slope of $d\alpha/d\ln Q$.  We take our reference
inputs at the $Z$ mass ($M_Z=91.188$~GeV) from global fits to
precision data, $\alpha_{EM}=1/128.9$ $\as=0.1185$, and
$\sin^2\theta_W=0.2312$~\cite{Eidelman:2004wy}. 

If the SM remains valid, at energies above the top mass, the couplings will
evolve with $\beta$-functions,
\bq
\label{eq:beta-SM}
\beta_i \; = \; \left\{ \frac{41}{6}, -\frac{19}{6}, -7 \right\} \; ,
\eq
while in the MSSM, above the masses of all the superpartners they are
\bq\label{eq:beta-MSSM}
\beta_i \; = \; \left\{ 11, 1, -3 \right\} \; .
\eq
Note the change in sign of $\beta_2$, which takes the $SU(2)$ gauge
coupling from being asymptotically free to asymptotically enslaved, an
indication of a {\em qualitative} difference between the SM and MSSM
at high energies.  The MSSM predictions for the gauge couplings are
model-dependent in that they are sensitive to the mass spectrum, and
thus the mechanism by which supersymmetry breaking is communicated to
the superpartners of SM fields.  We will assume that the electroweakly
interacting superpartners (charginos, neutralinos, sleptons, and Higgs)
have degenerate masses of 120 GeV, safely above the LEP II and Tevatron
bounds \cite{run2}, and modify the coupling evolution from the SM to
include the SUSY states at that energy.
However, the Run~II
bounds on colored superpartners are already somewhat better than
this~\cite{run2}, so we will assume their masses are $500$~GeV, also beyond
the current limits\footnote{This pattern of lighter weakly interacting
superpartners compared to heavier colored superpartners is also
common to most popular mechanisms of supersymmetry breaking, and is
largely induced by the renormalization group evolution of the
masses.}.  This implies that the electroweak couplings have a
threshold at 120 GeV from the electroweak gauginos, sleptons, and Higgses; one
at the top mass from the top quark;
and one at 500 GeV from the squarks, whereas the shift in
the evolution of the $SU(3)$ coupling away from the SM begins at 500 GeV.

%%%%%%%%%%%%%%%%%%%%%%%%%%%%%%%%%%%%%%%%%%%%%%%%%%%%%%%%%%%%%%%%%%%%%%%%

\section{Cross sections at the LHC
\label{sec:xsecs}}

Ideally, one would like to observe purely QED, weak and QCD
interactions to separately measure their gauge couplings.  In
practice, this is quite difficult.  For example, diphoton production
is on the order of only 10~fb for invariant masses above 1~TeV, and
more than an order of magnitude smaller for 2~TeV, where the lever arm
from running approaches usable size.  This gives far too few events to
perform a measurement.  Similarly, purely QCD processes suffer the
worst systematic uncertainties, mostly due to detector effects but
also on the theoretical side.

However, there is a purely weak process which is very easy to observe
and experimentally ``clean'': $pp\to W^*\to\ell\nu$, $\ell=e,\mu$.
This manifests itself as a highly energetic (``hard'') charged lepton
and missing energy in the transverse direction; one cannot reconstruct
the center-of-mass frame as the longitudinal momentum information of
the neutrino is lost.  The downside of this final state is that only
the transverse mass ($M_T$) can be measured, not the true invariant
mass, although they do track each other reasonably well.  We make
theoretical predictions for the transverse mass distribution using a
running coupling to calculate the differential cross section.  After
convolution with simulated detector effects, it could then be compared
with the data distribution.  Being forced to use the transverse mass
will weaken the measurement slightly, but not precipitously.

Drell-Yan charged lepton pairs, $pp\to Z/\gamma^*\to\ell^+\ell^-$,
allow for complete reconstruction of the kinematics and thus determination
of the invariant mass, but occur at a much smaller rate than $W^*$
events.  In addition, the $Z$ coupling to fermions is an admixture of
QED and the weak sector.  Nevertheless, it could potentially add
weight to the weak sector measurement via $W^*$.

To measure the QCD gauge coupling we choose a mixed QCD--QED process
out of necessity: gamma--jet production, $pp\to\aj$.  QCD jet
energies are not particularly
well-measured by LHC detectors, making comparison with theory of
energy-scale observables more difficult due to scale uncertainty.  By
requiring the QCD jet to be balanced by a photon, perhaps the worst
detector effects can be mitigated.  We discuss this issue further in
the relevant subsection below.

Our proposed relative gauge coupling measurement is thus a
deconvolution of the altered running on one pure- and two
mixed-coupling rates, with varying associated uncertainties.  We
describe in the following subsections the individual channels and
their likely expected uncertainties.  
In all cases, we rely on leading order estimates for the
rates (provided by {\sc MADGRAPH}~\cite{Stelzer:1994ta}, with the only
improvement coming from the running coupling.  While this is probably
not sufficient to match an experimental result with theory to the
desired accuracy, it should be sufficient to provide an estimate of
the statistical uncertainties, and we discuss systematic and
theoretical uncertainties below.  We use the CTEQ6L1 LO parton
distribution functions (PDFs)~\cite{Pumplin:2002vw} for our LO
calculations.  We present results assuming a combination of two
detectors, each of which will collect 300~fb$^{-1}$ at the LHC, and
3000~fb$^{-1}$ at the SLHC, running at $\sqrt{s}=14$~TeV center of
mass proton-on-proton collisions.

In all cases, the dominant uncertainties come from our knowledge of
the PDFs (primarily the gluon), higher-order QCD and electroweak
corrections, the integrated collider luminosity associated with the
data sample, and reconstruction of the event kinematics from
detector observables.  Fortunately for the Drell-Yan rates, the
higher-order QCD corrections are known to NNLO~\cite{DY-NNLO} and have
percent-level attendant uncertainties.  The gluon PDF is the least
well-known, especially at the large values of Feynman $x$ for
multi-TeV partonic collisions --- on the order of $10\%$.  Various
measurements can perhaps improve this at LHC, for example by DY itself
at lower invariant mass~\cite{Alekhin:2006zm}, but we propose to
eliminate some of the PDF (and all of the luminosity) uncertainty by
measuring {\it not absolute rates, but rates at large invariant mass
relative to those at lower invariant mass}, below the
scale where the MSSM content significantly 
alters gauge coupling running.  Thus, our
proposal is to measure the gauge couplings at high energy relative to
those known at or close to the $Z$ pole.

%%%%%%%%%%%%%%%%%%%%%%%%%%%%%%%%%%%%%%%%%%%%%%%%%%%%%%%%%%%%%%%%%%%%%%%%

\subsection{Drell-Yan {\boldmath $\ell\nu$} Pair Production
\label{sub:lv}}

\begin{figure}[t!]
\begin{center}
\includegraphics[scale=0.4]{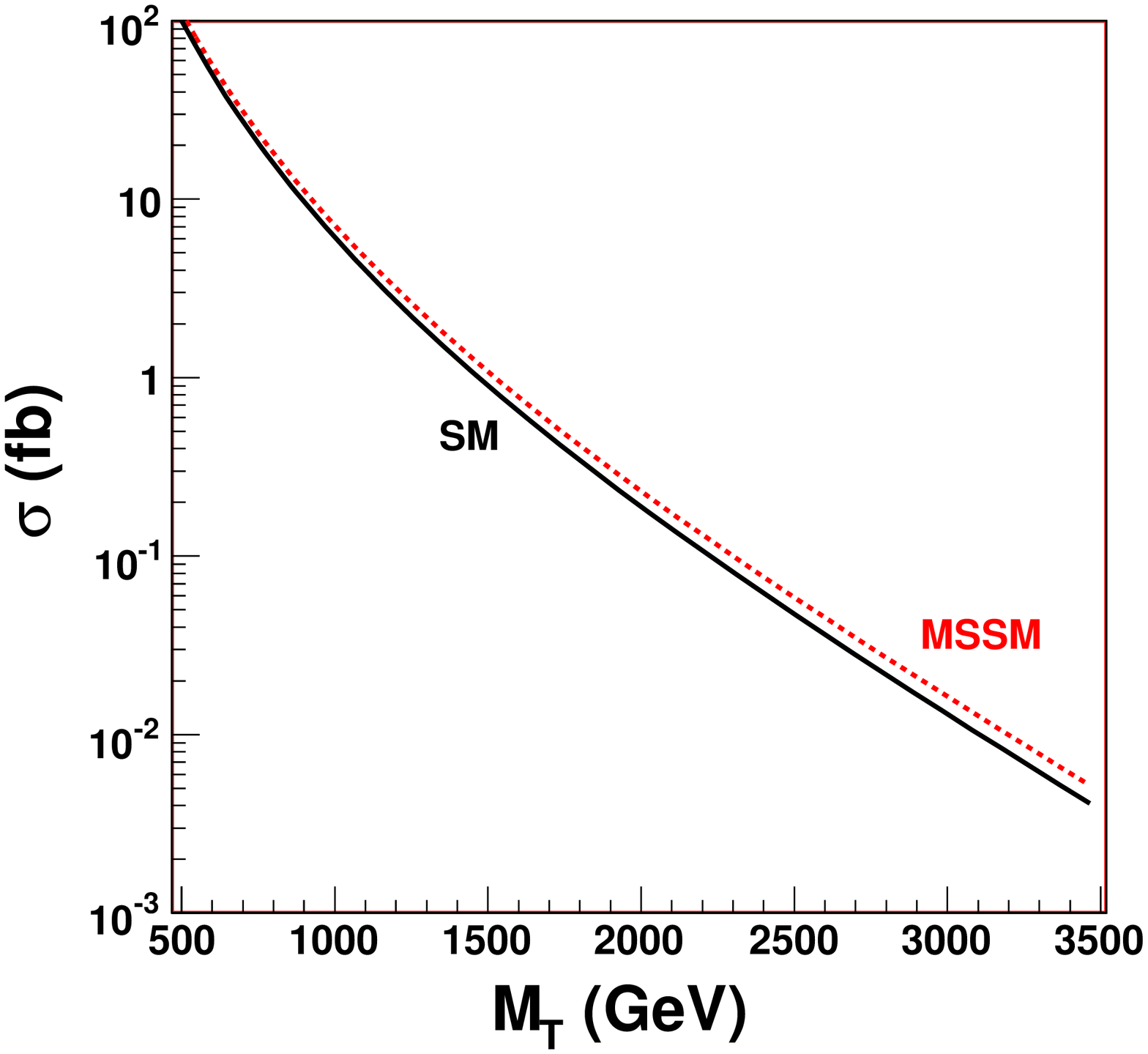}
\includegraphics[scale=0.4]{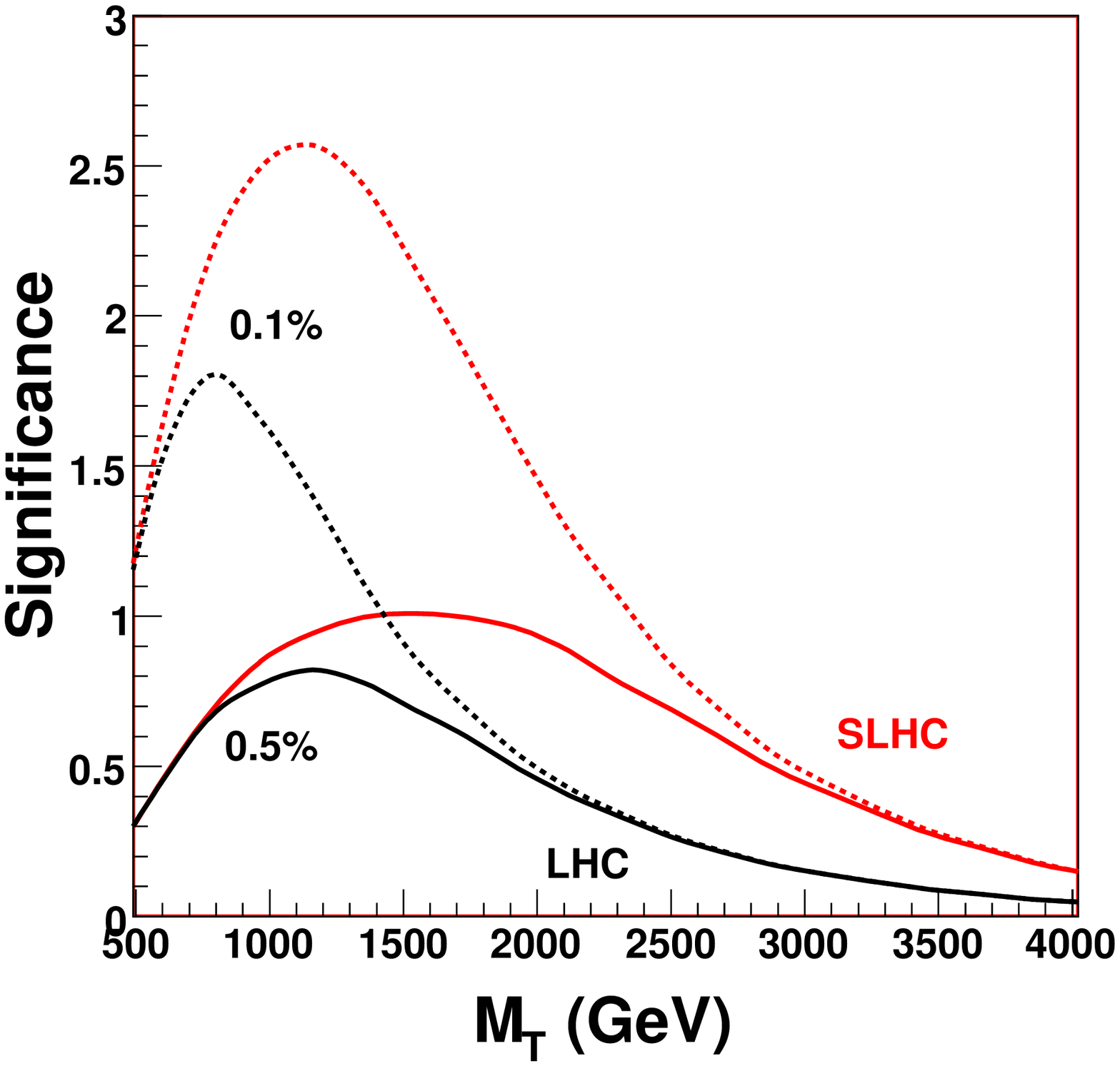}
\end{center}
\vspace{-10mm}
\caption{On the left is the rate of $pp\to\ell\nu$ assuming SM gauge
coupling evolution (solid curve) and MSSM evolution (dashed curve) as
a function of the cut on the minimum of the transverse mass.  On the
right is the significance as a function of a cut on the transverse
mass (including systematic uncertainties explained in the text) for
the LHC (lower curves) and SLHC (upper curves). 
The solid curves assume an electron energy resolution of $0.5\%$, whereas
the dashed curves assume $0.1\%$.}
\label{fig:dyw}
\end{figure}

Drell-Yan production of one charged and one neutral lepton takes place
at tree level through the process $q\bar{q}\to W^*\to\ell\nu$.  The
missing neutrino implies that the invariant mass of the $W$ cannot be
uniquely reconstructed at the LHC, but the transverse mass still
correlates well with the $W$ off-shellness, and shows the effect of
the different running coupling hypotheses at large transverse masses,
as shown in Figure~\ref{fig:dyw}.  We estimate our rates from LO
matrix elements; in a final analysis one should take advantage of NNLO
QCD \cite{DY-NNLO} and NLO EW \cite{Baur:2004ig} corrections, but
these effects are on the order of $10\%$ at the transverse masses of
interest, and thus will not affect our estimate of the statistical
analyzing power.  To simulate the acceptance of the detector, we
require the charged lepton to be central, with high $p_T$,
\bq
\label{eq:ellcut}
|y_{\ell} | \leq 2.5 \:, \qquad  \: p^{\ell}_T > 100~{\rm GeV} \, ,
\eq
(though in practice for such large transverse masses the $p_T$ cut is
irrelevant).  We apply an efficiency to identify a charged lepton of
$95\%$.  As we shall see below, the measurement of the lepton energy
is crucial to have a sensitivity, and we thus consider
only electrons, which can potentially be measured more
precisely, and not muons.

In order to estimate how well the large transverse mass $\ell \nu$
rate can be used to distinguish the Standard Model gauge couplings
from those in the MSSM, we assume that the low transverse mass region
has been matched to a SM calculation, which should remove the overall
luminosity uncertainty and the dominant PDF uncertainties.  We
estimate using the NLO inclusive rate~\cite{pavel} that the residual
PDF uncertainty should be of order $1\%$, and we assume, based on the
fact that NNLO calculations are available, that the residual
uncertainty from higher order, uncomputed QCD corrections is
negligible.  We assume a $0.5\%$ energy scale uncertainty for the
electron, and translate this back into a resultant uncertainty
in the cross section at that scale.  This assumption is optimistic, but
may be possible with enough collected luminosity.
Adding these uncertainties in quadrature
with the statistical uncertainty, we compute the significance of the
difference between SM and MSSM as a function of the cut on the minimum
of the transverse mass in Fig.~\ref{fig:dyw} for the LHC and SLHC.  We
see that the dominant limitation is from systematics, especially at
$M_T\sim 1$~TeV, and that for transverse mass cuts around $1-2$~TeV,
the SM and MSSM can be distinguished at about $1\sigma$ at
the (S)LHC.  In order to examine the importance of the electron energy
resolution, we also consider a $0.1\%$ energy scale uncertainty, and find
that this could raise the significance to $2\sigma$ or more.

One could imagine improving the significance by
considering bins of transverse mass.  The fact that the cross section
falls very rapidly with $M_T$ implies that a higher $M_T$ cut
sample is essentially uncorrelated with a lower $M_T$ cut sample, and
thus one could potentially make more than one measurement with similar
statistical significance in different bins of $M_T$, improving the overall
significance.  We leave such refinements to future work.

%

%%%%%%%%%%%%%%%%%%%%%%%%%%%%%%%%%%%%%%%%%%%%%%%%%%%%%%%%%%%%%%%%%%%%%%%%

\subsection{Drell-Yan {\boldmath $\ell^+\ell^-$} Production
\label{sub:ll}}

\begin{figure}[t!]
\begin{center}
\includegraphics[scale=0.4]{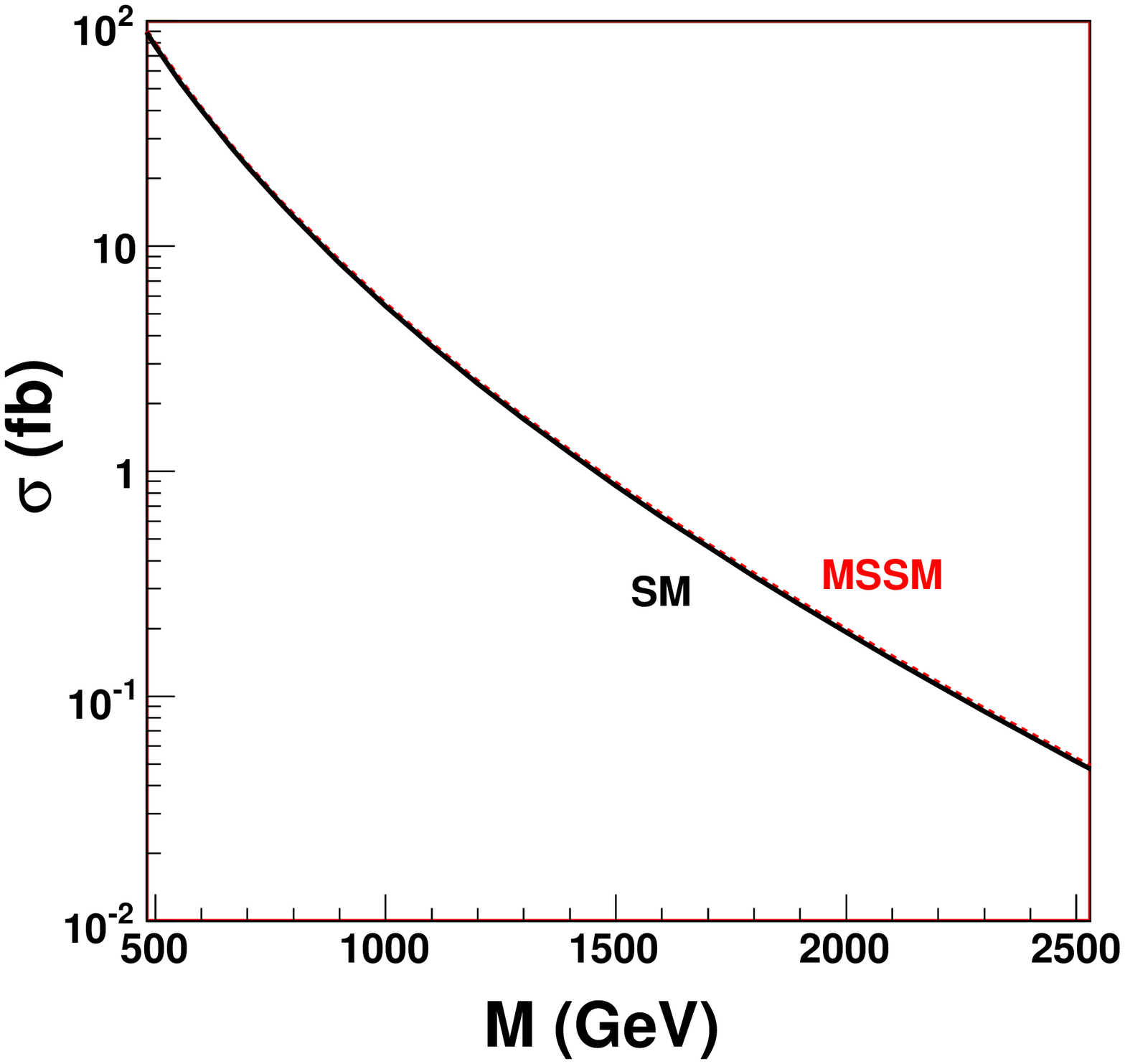}
\includegraphics[scale=0.4]{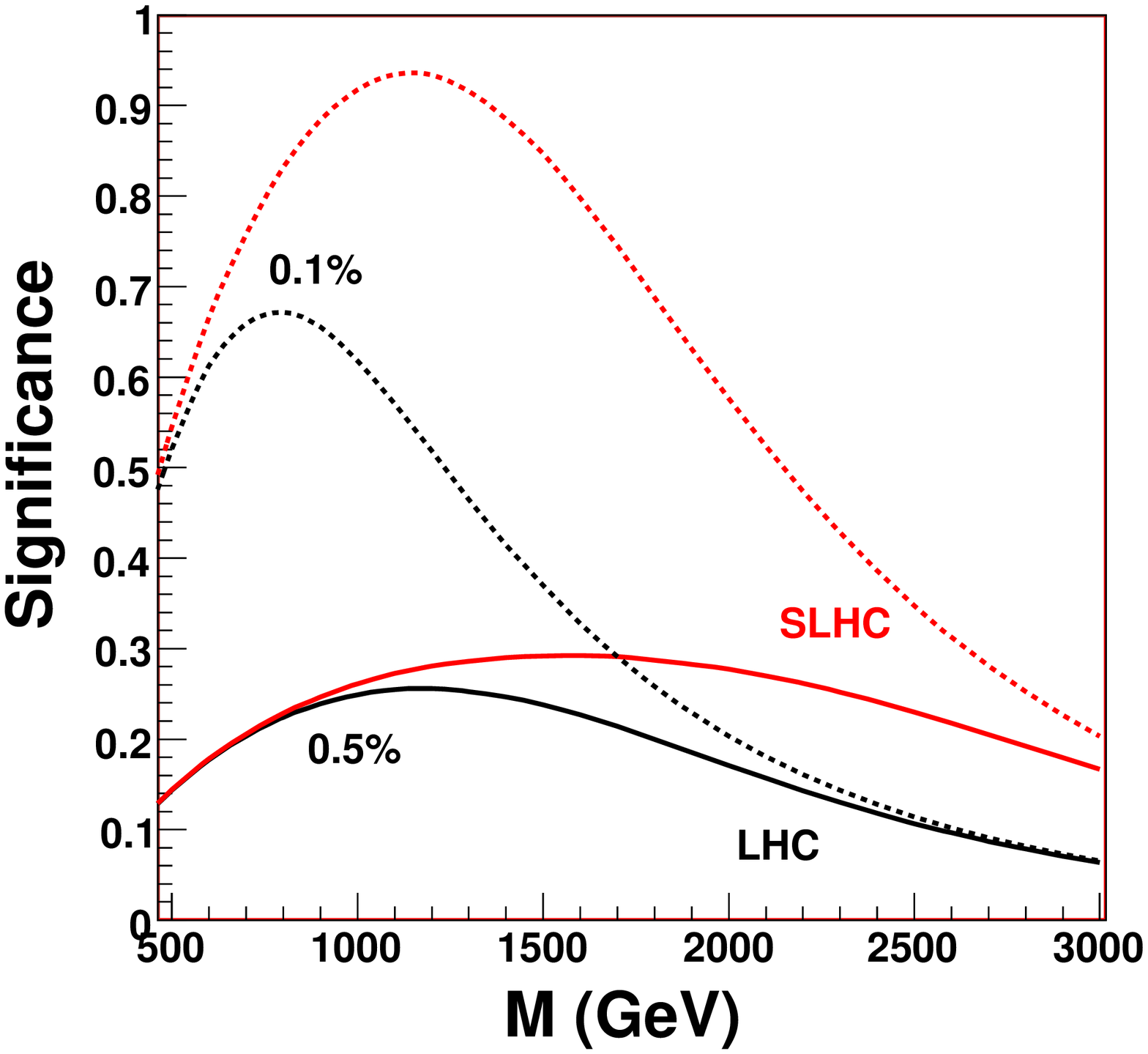}
\end{center}
\vspace{-10mm}
\caption{On the left is the rate of $pp\to\ell^+\ell^-$ assuming SM
gauge coupling evolution (solid curve) and MSSM evolution (dashed
curve) as a function of the cut on the minimum of the transverse mass.
On the right is the significance as a function of a cut on the
transverse mass (including systematic uncertainties explained in the
text) for the LHC (lower curves) and SLHC (upper curves).
The solid curves assume an electron energy resolution of $0.5\%$, whereas
the dashed curves assume $0.1\%$.}
\label{fig:dyz}
\end{figure}

Charged lepton pair production is mediated by both the photon and $Z$
boson.  Both couplings are a combination of the $SU(2)\times U(1)$
couplings $g_1$ and $g_2$, and in this case we are able to directly
reconstruct the virtuality of the bosons from the charged pair
invariant mass.  We once again restrict ourselves to electrons
and impose the same acceptance cuts on the leptons as
before, Eq.~(\ref{eq:ellcut}) with an identification efficiency of
$95\%$ per lepton, and once again assume a $0.5\%$ ($0.1\%$)
energy resolution for
(both) electrons, and $1\%$ PDF uncertainty.  The SM and MSSM
rates and MSSM versus SM significance are shown in Fig.~\ref{fig:dyz}, and
are considerably smaller than for the corresponding charged current
case.  This is partially because the effect of the running coupling is
less, but also because the process is more sensitive to the energy
resolution on the charged leptons, and the correpsonding uncertainty
thus larger.  Our conclusion is that this process is unlikely to be
useful in its own right, though it could perhaps be included in a more
global analysis based on either charged-neutral production or $\gamma$--jet 
production to add some information.  Again, the cross section
falls rapidly, and one could imagine a binned analysis with 
small correlation between bins and improved overall significance.

%

%%%%%%%%%%%%%%%%%%%%%%%%%%%%%%%%%%%%%%%%%%%%%%%%%%%%%%%%%%%%%%%%%%%%%%%%

\subsection{Photon-jet production
\label{sub:gamjet}}

The photon-jet production rate will be the least well-measured/predicted 
of any
that we propose, due to the initial-state gluon PDF at LO and the jet
and photon energy scale uncertainty (detector capability) in the
multi-TeV regime.  
However, these drawbacks can be balanced by a larger rate (thus smaller
statistical error), and sensitivity to $g_3$ which evolves faster than
$g_2$ or $g_1$.
Ideally, one would prefer to examine the dijet rate
at large invariant mass, because it is a purely QCD process, and also
has approximately twice the rate enhancement in the MSSM case compared
to the photon-jet cross section.  (This is because $\aem$ runs very
little compared to $\as$.)  However, the dijet rate suffers from twice
the gluon uncertainty, and prohibitively large jet energy uncertainties.

Photons, in contrast, tend to be much better-measured than jets,
because they shower electromagnetically in the detector and can be
calibrated by comparison with electrons, whose momentum can be
measured by their trajectory in the magnetic field as observed by the
tracking system.  (In fact, the jet energy scale is calibrated in
experiment using photons, in turn calibrated from electrons.)  Our
idea is that in high $p_T$ photon-jet events, one would assume that
the jet balances the photon in transverse momentum, then use the
well-measured flight direction of the jet to completely reconstruct
the system.

However, the dearth of multi-TeV electrons leads to the na\"ive
estimate that the photon energy scale in our region of interest will
not be better than about $3\%$, about the same as jets.  This would
immediately kill our proposal.  The solution is to rely on photons at
very high transverse momenta using the (well-known rate of)
electron-conversion photons, which provides a source of equally
high-energy electrons for ``self-calibration''~\cite{tom}.  This is
likely to allow for at least $1\%$ energy scale uncertainty, and
possibly $1/2\%$, over the lifetime of the first LHC run.  This could
be reduced by about a factor of about 3 (i.e. $\sqrt{10}$) using the
ten times statistics available at the LHC luminosity upgrade, the
SLHC~\cite{Gianotti:2002xx}.

Although the QCD NLO results are known for photon-jet production, and
are large at low invariant mass~\cite{gamjet-NLO}, the results for
higher invariant masses are not readily available.  We neglect
fragmentation processes, and do not apply a K-factor to our LO
results, which we believe to be a conservative approximation to the
available rate at LHC.  We also do not consider ``background'' from
dijet production, as the rate for misidentifying a jet as a photon
occurs at about the $10^{-4}$ level, while the cross section ratio is
more on the order of $10^2$--$10^3$.  Naturally these approximations
should be investigated further, when detector simulation is added.  To
simulate the detector acceptance, we require both the photon and the
jet to be central:
\bq\label{eq:cuts}
|\eta_j| < 4.0 \; , \qquad |\eta_{\gamma}| < 2.5 \; ;
\eq
and include an $80\%$ efficiency factor to identify a photon~\cite{ATLAS,CMS}.

\begin{figure}[ht!]
\begin{center}
\includegraphics[width=15cm]{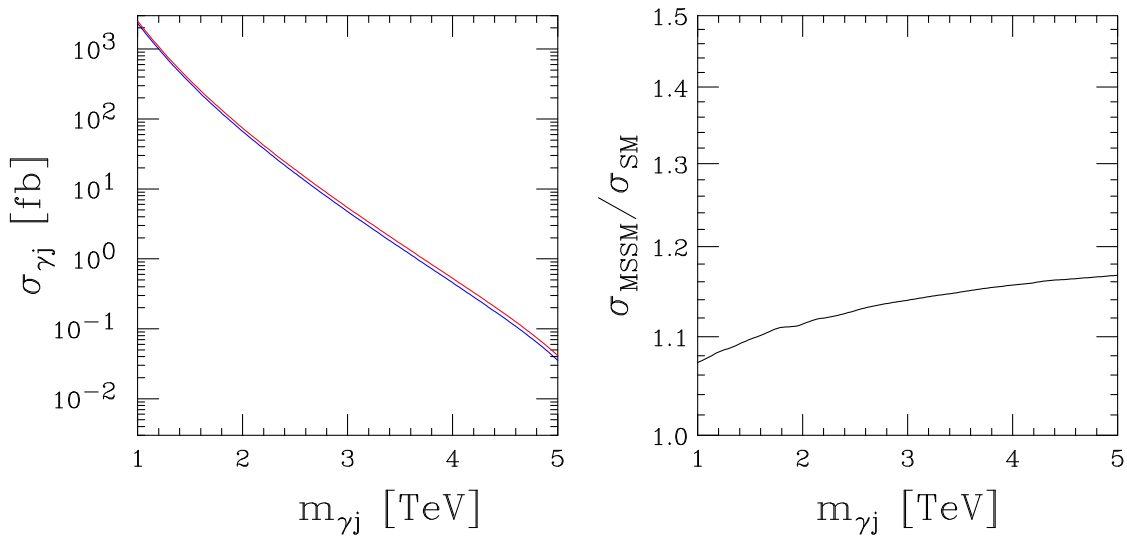}
\end{center}
\vspace{-6mm}
\caption{Left: the leading-order photon-jet cross section at the LHC,
in the SM (lower curve) and MSSM (upper curve) as a function of
photon-jet invariant mass.  The cross section falls off a little
slower than $m_{\aj}^4$.  Right: the ratio of MSSM to SM cross
sections at the LHC, also as a function of $m_{\aj}$.  The logarithmic
deviation due to the running coupling is readily apparent.}
\label{fig:xsec-gamjet}
\end{figure}
\begin{figure}[ht!]
\begin{center}
\includegraphics[width=15cm]{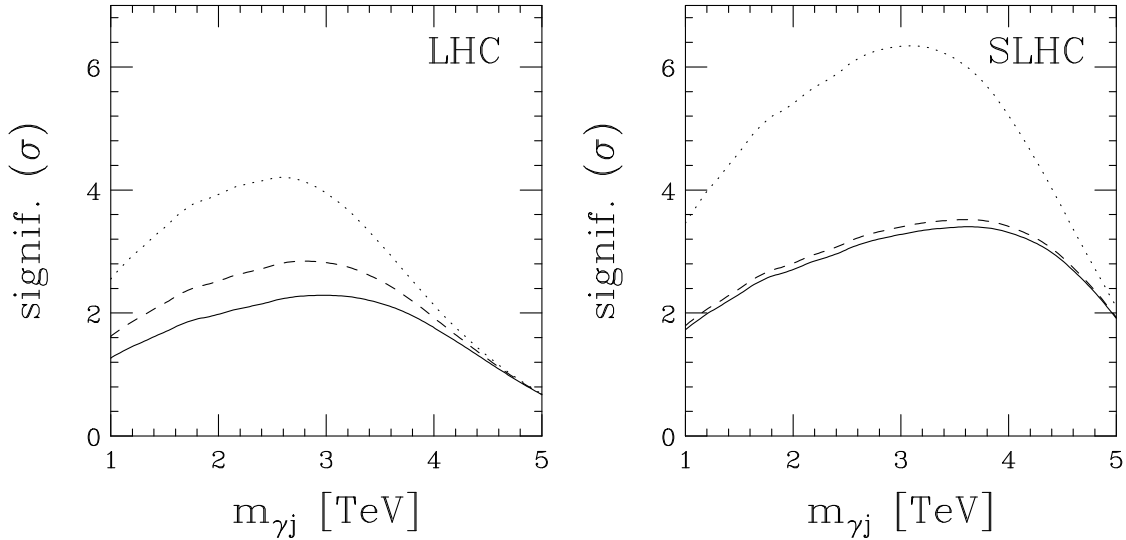}
\end{center}
\vspace{-6mm}
\caption{Results for the expected significance of MSSM-running gauge
couplings v. the SM at LHC (left) and SLHC (right).  The solid curve
at LHC is for the more pessimistic assumptions of $4\%$ relative PDF
uncertainty and $1\%$ photon energy scale uncertainty, which dashed is
for $0.5\%$ photon energy scale uncertainty, and dotted is further
reducing the relative PDF uncertainty to $2\%$.  The curves are the
same at SLHC, except that the corresponding photon energy scale
uncertainties are $\sqrt{10}$ smaller than those at the LHC, due to
the larger available sample sizes for calibration.  For two detectors,
the LHC assumes 300~fb$^{-1}$ each, and the SLHC ten times that.}
\label{fig:results-gamjet}
\end{figure}

We first show the LO cross section and MSSM/SM cross section ratio in
Fig.~\ref{fig:xsec-gamjet} as a function of photon-jet invariant mass.
The former gives one an idea of what event statistics will be
available at LHC, while the latter distinctly shows the dominant
logarithmic behavior of including the MSSM spectrum in the gauge
coupling running.  Approximately $85\%$ of the running is due to the
evolution of $\alpha_s$, the remaining part due to the QED coupling.

Our results are summarized in Fig.~\ref{fig:results-gamjet} for the
LHC and SLHC.  The curves represent the signal significance, i.e. the
deviation of the MSSM rate from the expected SM rate.  Keep in mind
this will ultimately be done as a {\it relative} measurement of the
rate at large invariant masses to that at around 200~GeV.  The
significance is calculated as the MSSM--SM rate separation in terms of
SM rate uncertainty, adding in quadrature the uncertainties from
statistics, photon energy scale and PDFs (dominantly the gluon,
although the high-$x$ quark uncertainties would also have to be known
more precisely than at present).  At the moment, high-$x$ gluons are
known to only $10-15\%$ in our region of interest; quite poorly.  How
much uncertainty exists in the relative measurement at high invariant
mass to low is not well-studied~\cite{pavel,frank}.  However, it should
be possible to improve this using Drell-Yan lepton data over a range
of invariant mass and rapidity to tighten up the uncertainty bands
considerably~\cite{frank}.  This is possible in part because the QCD
theory uncertainty on differential DY is negligible~\cite{DY-NNLO}.
To illustrate the range of potential, we first start with a
pessimistic but reasonable $4\%$ PDF uncertainty at high $x$, and
$1\%$ on the photon energy normalization.  This is shown by the solid
lines.  The dashed lines are the result of reducing the photon energy
scale uncertainty by a factor two, to half a percent.  The dotted
curves are additionally a reduction of the PDF uncertainty by a factor
two, from $4\%$ to $2\%$.  This represents the most optimistic
scenario.

It is clear that at the LHC the limiting factor is the PDF uncertainty
at high $x$, dominantly the gluon, although improving the photon
energy scale calibration results in a significant increase of
analysing power, from about a $2\sigma$ observation to almost
$3\sigma$.  LHC could achieve $4\sigma$ if the convoluted PDFs at high
$x$ can be brought under control to the $2\%$ level, relative to those
as lower $x$ where the lower invariant mass cross section is made,
which is not a totally unreasonable assumption.  At the SLHC, the much
higher event rate available for photon energy scale calibration
largely removes that as a large source of uncertainty.  Instead, the
gluon PDF dominates.  If a $2\%$ uncertainty can be achieved, not only
would this measurement result in a beautiful observation of altered
running of the gauge couplings, it would allow for a first measurement
of the actual gauge coupling values (primarily $\alpha_s$), somewhere
in the $20\%$ range at SLHC.

%%%%%%%%%%%%%%%%%%%%%%%%%%%%%%%%%%%%%%%%%%%%%%%%%%%%%%%%%%%%%%%%%%%%%%%%

\section{Discussion and conclusions
\label{sec:conc}}

We propose using three different high momentum-transfer scattering
processes at LHC, Drell-Yan lepton pairs (charged and mixed
charged-neutral) and photon-jet events, to study the running of the SM
gauge couplings at high energy.  Our motivation is the well-known
observation that the gauge couplings appear to almost converge at
about $10^{15}$~GeV, and are predicted to do so in many models of new
physics that respect a GUT symmetry at that scale.  The generic
prediction is additional particle content at or near the electroweak
scale which alters the gauge couplings' evolution such that they
converge at a single energy scale, at which point they become one
coupling.  Our proposal suggests that one use these processes as a
probe of the GUT hypothesis, and we show that in the specific case of
the MSSM, we are able to distinguish the MSSM from the SM using the
evolution of the gauge couplings alone.  While this is not in itself a
proof of unification of couplings, it is at least suggestive of the
first step.  Further, the combination of this indirect effect of the
new physics with direct detection of some of the spectrum can reveal
states which may be difficult to see directly, and provides a measure
of the sum of the elecroweak charges active in the evolution, which
may be obscured by electroweak symmetry breaking in direct
observation.

We have examined three processes involving Standard Model external states
at high momentum transfer,
and have shown that we find a potentially very significant difference
between SM and MSSM extrapolations to high $Q^2$ in the case of
the semi-weak production of a jet and a photon, moderately significant
differences in the charged DY production of an electron and a neutrino, and
not very significant differences in the DY $e^+ e^-$ channel.  We have
made optimistic but not impossible assumptions about the potential to
measure jet and electron energies at the LHC detectors, and have shown
that in each case this is the limiting factor to realize the idea.

Our results demonstrate that the (S)LHC is potentially more of a
precision measurement machine than is normally credited.  For these
measurements, and prospects for adding other channels that are purely
QCD, the limiting factors are likely to be the gluon PDF uncertainty at
large values of Feynman $x$, and the precise measurement of energies
of jets, photons, and leptons.  We have chosen optimistic (but
hopefully achievable) estimates of the detector capabilities and PDF
uncertainties, and the results are tantalizing.  They provide strong
encouragement to the continued heroic effort toward reducing these
uncertainties via other measurements and creative experimental
and theoretical techniques, both
those ongoing at Tevatron and those that can be performed over the
lifetime of LHC.

%%%%%%%%%%%%%%%%%%%%%%%%%%%%%%%%%%%%%%%%%%%%%%%%%%%%%%%%%%%%%%%%%%%%%%%%

\subsection*{Acknowledgments}

We would like to thank Pavel Nadolsky and Frank Petriello for
discussions about PDF uncertainties, and Tom LeCompte for numerous
discussions about detector capabilities at multi-TeV energies and for
keeping us within the realm of the technically feasible.  This
research was supported in part by the U.S. Department of Energy under
grant No. DE-FG02-91ER40685 (DR) and contract No. DE-AC02-06CH11357
(TT).

%%%%%%%%%%%%%%%%%%%%%%%%%%%%%%%%%%%%%%%%%%%%%%%%%%%%%%%%%%%%%%%%%%%%%%%%
%%% References
%%%%%%%%%%%%%%%%%%%%%%%%%%%%%%%%%%%%%%%%%%%%%%%%%%%%%%%%%%%%%%%%%%%%%%%%


\baselineskip15pt

\begin{thebibliography}{19}

%\bibitem{SUSY}
%For reviews of supersymmetry, see:
%\bibitem{Martin:1997ns}
%  S.~P.~Martin,
  %``A supersymmetry primer,''
%  arXiv:hep-ph/9709356; \\
  %%CITATION = HEP-PH 9709356;%%
%\bibitem{Aitchison:2005cf}
%  I.~J.~R.~Aitchison,
  %``Supersymmetry and the MSSM: An elementary introduction,''
%  arXiv:hep-ph/0505105.
  %%CITATION = HEP-PH 0505105;%%

\bibitem{Bourilkov:2006rz}
  D.~Bourilkov,
  %``Gauge coupling unification, SUSY scale and strong coupling running,''
  AIP Conf.\ Proc.\  {\bf 842}, 634 (2006).
  %[arXiv:hep-ph/0602168].
  %%CITATION = HEP-PH 0602168;%%

\bibitem{Vlike-quarks}
%\bibitem{Choudhury:2001hs}
  D.~Choudhury, T.~M.~P.~Tait and C.~E.~M.~Wagner,
  %``Beautiful mirrors and precision electroweak data,''
  Phys.\ Rev.\ D {\bf 65}, 053002 (2002); \\
  %[arXiv:hep-ph/0109097];
  %%CITATION = HEP-PH 0109097;%%
%\bibitem{Morrissey:2003sc}
  D.~E.~Morrissey and C.~E.~M.~Wagner,
  %``Beautiful mirrors, unification of couplings and collider phenomenology,''
  Phys.\ Rev.\ D {\bf 69}, 053001 (2004); \\
  %[arXiv:hep-ph/0308001];
  %%CITATION = HEP-PH 0308001;%%
%\bibitem{Giudice:2004tc}
  G.~F.~Giudice and A.~Romanino,
  %``Split supersymmetry,''
  Nucl.\ Phys.\ B {\bf 699}, 65 (2004) \\
  {[Erratum-ibid.\ B {\bf 706}, 65 (2005)]}.
  %[arXiv:hep-ph/0406088].
  %%CITATION = HEP-PH 0406088;%%

\bibitem{run2}
  See \eg: \\
  {\tt http://www-cdf.fnal.gov/physics/exotic/exotic.html} for CDF and \\
  {\tt http://www-d0.fnal.gov/public/new/new\_public.html} for D\O\ .

\bibitem{Dawson:1983fw}
  S.~Dawson, E.~Eichten and C.~Quigg,
  %``Search For Supersymmetric Particles In Hadron - Hadron Collisions,''
  Phys.\ Rev.\ D {\bf 31}, 1581 (1985).
  %%CITATION = PHRVA,D31,1581;%%

\bibitem{Eidelman:2004wy}
  S.~Eidelman {\it et al.}  [Particle Data Group],
  %``Review of particle physics,''
  Phys.\ Lett.\ B {\bf 592}, 1 (2004).
  %%CITATION = PHLTA,B592,1;%%

\bibitem{Stelzer:1994ta}
  T.~Stelzer, F.~Long,
  %%''AUTOMATIC GENERATION OF TREE LEVEL HELICITY AMPLITUDES.''
  Comput.{} Phys.{} Commun.{} \textbf{81} (1994) 357.
  %[arXiv:hep-ph/9401258].
  %%CITATION = hep-ph/9401258;%%

\bibitem{Pumplin:2002vw}
  J.~Pumplin {\it et al.},
  %D.~R.~Stump, J.~Huston, H.~L.~Lai, P.~Nadolsky and W.~K.~Tung,
  %``New generation of parton distributions with uncertainties from
  % global QCD analysis,''
  JHEP {\bf 0207}, 012 (2002)
  %[arXiv:hep-ph/0201195].
  %%CITATION = HEP-PH 0201195;%%

\bibitem{DY-NNLO}
%\bibitem{Anastasiou:2003ds}
  C.~Anastasiou, L.~J.~Dixon, K.~Melnikov and F.~Petriello,
  %``High-precision QCD at hadron colliders: Electroweak gauge boson
  %rapidity distributions at NNLO,''
  Phys.\ Rev.\ D {\bf 69}, 094008 (2004); \\
  %[arXiv:hep-ph/0312266].
  %%CITATION = HEP-PH 0312266;%%
%\bibitem{Melnikov:2006di}
  K.~Melnikov and F.~Petriello,
  %``The W boson production cross section at the LHC through O(alpha(s)**2),''
  Phys.\ Rev.\ Lett.\  {\bf 96}, 231803 (2006); \\
  %[arXiv:hep-ph/0603182].
  %%CITATION = HEP-PH 0603182;%%
%\bibitem{Melnikov:2006kv}
  K.~Melnikov and F.~Petriello,
  %``Electroweak gauge boson production at hadron colliders through
  %O(alpha(s)**2),''
  arXiv:hep-ph/0609070.
  %%CITATION = HEP-PH 0609070;%%

\bibitem{Alekhin:2006zm}
  S.~Alekhin, K.~Melnikov and F.~Petriello,
  %``Fixed target Drell-Yan data and NNLO QCD fits of parton distribution
  %functions,''
  arXiv:hep-ph/0606237.
  %%CITATION = HEP-PH 0606237;%%

\bibitem{Baur:2004ig}
  U.~Baur and D.~Wackeroth,
  %``Electroweak radiative corrections to p (anti-)p --> W+- --> l+- nu
  %beyond the pole approximation,''
  Phys.\ Rev.\ D {\bf 70}, 073015 (2004).
  %[arXiv:hep-ph/0405191].
  %%CITATION = HEP-PH 0405191;%%

\bibitem{ATLAS}
  ATLAS TDR, report CERN/LHCC/1999-15 (1999).

\bibitem{CMS}
  CMS TDR, report CERN/LHCC/2006-001 (2006).

\bibitem{tom}
  Tom LeCompte, private communication.

\bibitem{Gianotti:2002xx}
  F.~Gianotti \etal,
  %``Physics potential and experimental challenges of the LHC luminosity
  %upgrade,''
  Eur.\ Phys.\ J.\ C {\bf 39}, 293 (2004).
  %[arXiv:hep-ph/0204087].
  %%CITATION = HEP-PH 0204087;%%

\bibitem{gamjet-NLO}
%\bibitem{Catani:2002ny}
  S.~Catani, M.~Fontannaz, J.~P.~Guillet and E.~Pilon,
  %``Cross section of isolated prompt photons in hadron hadron collisions,''
  JHEP {\bf 0205}, 028 (2002); \\
  %[arXiv:hep-ph/0204023].
  %%CITATION = HEP-PH 0204023;%%
%\bibitem{Aurenche:2006vj}
  P.~Aurenche, M.~Fontannaz, J.~P.~Guillet, E.~Pilon and M.~Werlen,
  %``A new critical study of photon production in hadronic collisions,''
  arXiv:hep-ph/0602133.
  %%CITATION = HEP-PH 0602133;%%

\bibitem{pavel}
  %\cite{Balazs:1997xd}
%\bibitem{Balazs:1997xd}
  C.~Balazs and C.~P.~Yuan,
  %``Soft gluon effects on lepton pairs at hadron colliders,''
  Phys.\ Rev.\ D {\bf 56}, 5558 (1997). \\
  %[arXiv:hep-ph/9704258];
  %%CITATION = HEP-PH 9704258;%%
  We are grateful to Pavel Nadolsky, for help in extracting this result
from the RESBOS code.

\bibitem{frank}
  Frank Petriello, private communication based on Ref.~\cite{DY-NNLO}.

\end{thebibliography}
\end{document}